\baselineskip=12 pt
\font\bigbf=cmbx10 scaled 1300

\newcount\eqnumber
\def\clreqnumber{\global\eqnumber=0} \clreqnumber
\def\EQN#1#2$${\global\advance\eqnumber by1%
  \eqno\hbox{\rm({}\the\eqnumber#1)}$$\def\name{#2}\ifx\name\empty%
  \else\xdef#2{({}\the\eqnumber#1)\noexpand}\fi\ignorespaces}
\def\eqn#1$${\EQN{}{#1}$$}
\def\eqna#1$${\EQN{a}{#1}$$}
\def\eqnb#1$${\global\advance\eqnumber by-1 \EQN{b}{#1}$$}
\def\eqnc#1$${\global\advance\eqnumber by-1 \EQN{c}{#1}$$}
\def\lasteqn{{\rm({}\the\eqnumber)}}
\def\nexteqn{\advance\eqnumber by1 {\rm({}\the\eqnumber)}\advance
  \eqnumber by-1 }

\centerline{\bigbf Dissipation for Euler's Disk and a Desktop }
\centerline{\bigbf Demonstration of Coalescing Neutron Stars}

\bigskip

\noindent
{\bf Lars Bildsten}

\bigskip
\noindent
{\it Institute for Theoretical Physics, Kohn Hall, 
University of California, Santa Barbara, California 93106, USA; 
e-mail: bildsten@itp.ucsb.edu}

\bigskip

\centerline{Submitted to American Journal of Physics, Feb 11, 2001}

\bigskip

 I show that the recent calculation of Moffatt's regarding the viscous
dissipation of a spinning coin overlooked the importance of the
finite width of the viscous boundary layer. My new estimates are more
  in accord with that observed. I also point out that the frequency
  ``chirp'' of the specially designed toy ``Euler's Disk'' is 
  similar to that expected during the last few minutes of the life of
  a coalescing binary of two neutron stars. As such, this toy is an
  excellent desktop demonstration for the expected phenomena.

\bigskip

\noindent
{\bf I. INTRODUCTION}

\bigskip
  
   It is natural to wonder where the energy is dissipated that allows a
spinning, wobbling, coin to eventually come to rest.  Moffatt$^1$
showed that a source of damping for ``Euler's disk'' (see
http://www.eulersdisk.com for information on this toy) is viscous
dissipation in the air flowing between the disk (of radius $a=3.75 \
{\rm cm}$) and the base it is oscillating upon at frequency $\Omega$
(see Figure 1 of Moffatt 2000). I show here that Moffatt
underestimated this dissipation by neglecting the finite time it takes
for viscosity to act across this thin layer.

\bigskip
\noindent 
{\bf II. VISCOUS DISSIPATION IN THE PRESENCE OF A BOUNDARY LAYER}
\bigskip

  For small angles $\alpha$ between the base and disk, the frequency
of wobble is $\Omega^2=4g/a\alpha$, where $g=980 \ {\rm cm^2 \
s^{-1}}$ is the Earth's gravitational acceleration.  Moffatt assumed
that the viscous boundary layer extended completely across the
vertical gap (of width $\approx a\alpha$) between the disk and the
base. For this to be true, the time for the viscous flow (with
kinematic viscosity$^2$ $\nu\equiv\mu/\rho\approx 0.15 \ {\rm cm^2 \
s^{-1}}$ at 20C) to be established across the gap, $t_v \approx
(a\alpha)^2/\nu$, must be shorter than the oscillation period; $\Omega
t_v \ll 1$.  Moffatt's calculation is only valid for angles
smaller than
$$
\alpha_c\approx\left(\nu^2\over 4ga^3\right)^{1/3}\approx 4.8\times
10^{-3} \ {\rm rad}\approx 0.3 \ {\rm deg},\eqn\alphac
$$
which corresponds to oscillating frequencies higher than $\approx 70 \
{\rm Hz}$, or just the very end of the collapse. For most larger
angles, $\alpha > \alpha_c$, the viscous dissipation occurs in the
oscillating thin boundary layers$^3$ at the disk and base surfaces of
width $\delta\approx (2\nu/\Omega)^{1/2}\ll \alpha a$.
The dissipation rate of the air flowing at speed
$u\approx\Omega a $ is then roughly 
$$
\Phi\sim \mu a^2 \delta \left(du\over dz\right)^2\sim {4g\mu a^3\over \alpha
\delta}\propto {1\over \alpha^{5/4}},\eqn\phia
$$
which is {\it larger} than Moffatt's value by
 $\alpha a/\delta\approx(\alpha/\alpha_c)^{3/4}\gg 1 $. Just as
viscous flow is more difficult in a smaller pipe, forcing the
viscous boundary layer into a thin layer increases the
dissipation. Accurately calculating the dimensionless prefactors for
equation (2) and the transition to Moffatt's scalings for $\Phi$ when
$\alpha<\alpha_c$ requires a full solution of the oscillating viscous
boundary layer flow in the disk geometry.

However, the parameter dependences of $\Phi$ are accurate when $\alpha
> \alpha_c$ and allow us to find the scalings of $\Omega$ and $\alpha$
with time as the toy ``spins-up'' and falls to the base  under this
sole dissipation mechanism (see recent discussions$^{4-5}$ of 
surface friction). The energy
of the disk (of mass $M$) is $E=3Mga\alpha/2$, and using $\Phi$ from
equation (2) in $-\Phi=dE/dt$, the equation that relates the final
angle ($\alpha_f$) to the initial ($\alpha_i$) is
$$
\alpha_f^{9/4}=\alpha_i^{9/4}-{t_f-t_i\over t_{\rm BL}},\eqn\alpaha
$$
where $t_i$ and $t_f$ are the initial and final times,  and 
$$
t_{\rm BL}={M\over 6\mu a}\left(\nu^2\over a^3g\right)^{1/4}={M\over 6\mu
a}(4\alpha_c^3)^{1/4}\approx 42 \ {\rm min},\eqn\alphab
$$
for the designed toy with $M=400 \ {\rm g}$.  The time to collapse to
$\alpha \ll \alpha_i$ is $t_c=t_{\rm
BL}\alpha_i^{9/4}$, which gives $\approx 50$ seconds from an initial angle of
ten degrees, when the oscillating frequency is $\approx 12$ Hz. 
This roughly agrees with that observed, 
though, as earlier emphasized, the parameter dependences of equation
(2) are more reliable than the numerical prefactors. They should be
tested, and if true, could be used to ``engineer'' a better toy. 
An immediate experimental check is the predicted scaling with time; 
$\alpha(t)=[(t_c-t)/t_{\rm BL}]^{4/9}$. 
For $\alpha > \alpha_c$ the frequency should scale as 
$$\Omega\propto (t_c-t)^{-2/9}, \eqn\omehas
$$
different than Moffatt's scaling, which 
had an exponent of $-1/6$.

 All of my discussion and calculations to this point have assumed that
the air-flow remains laminar in the oscillating boundary layer. This
appears to be a fair assumption for the designed toy.  Turbulent flow
could ensue if the Reynolds number
$$
{\rm Re}\approx {a^2\Omega\over \nu}\approx {1\over
\alpha_c^2}\left(\alpha_c\over \alpha\right)^{1/2}\approx 4\times 10^4 
\left(\alpha_c\over \alpha\right)^{1/2}, \eqn\omj
$$
becomes too large. In conventional pipe flow$^3$ the onset of
turbulence occurs when ${\rm Re}$ exceeds 6000, while in oscillating
viscous boundary layers$^6$, the onset only occurs for 
${\rm Re}>10^5$. The viscous flow for this toy appears
to be close to the onset of turbulence when $\alpha\sim\alpha_c$, but
before that small angle, the laminar calculations should remain
valid. 

\bigskip

\noindent 
{\bf III. COMPARISON TO A COALESCING NEUTRON STAR BINARY}

\bigskip

 The frequency sweep and time to collapse of Euler's Disk is similar
enough to the final gravitational wave output of two coalescing
neutron stars that it makes an excellent desktop demonstration of this
phenomena. The gravitational wave chirp from the two neutron stars in
a tight binary sweeps from 20 Hz to 1000 Hz in a few minutes$^7$ and
will likely be detected by gravitational wave interferometers
(e.g. the Laser Interferometer Gravitational Wave Observatory, LIGO) within
ten years. The orbital energy is $\propto - \Omega^{2/3}$ ($\Omega$ is
the orbital frequency in this context) and decreases at a rate
$\propto \Omega^{10/3}$ set by gravitational wave emission. The
finite time singularity for the orbital frequency is $\Omega\propto
(t_c-t)^{-3/8}$, steeper than equation (5). The maximum frequency is
fixed by the tidal destruction of both neutron stars, a catclysmic event
that many attribute as the energy source for gamma ray bursts.

\bigskip

\noindent 
$^1$  Moffatt, H. K., ``Euler's disk and its finite time singularity,'' 
Nature, {\bf 404}, 833-834 (2000). 

\noindent
$^2$ Denny, M. W., {\it Air and Water: The Biology and Physics of Life's
Media} (Princeton, Princeton Univ. Press, 1993) 

\noindent
$^3$ Landau, L. D. \& Lifshitz, E. M. {\it Fluid Mechanics}
(Oxford: Pergamon Press, 1959)

\noindent
$^4$ van den Engh, G., Nelson, P., Roach, J., ``Numismatic gyrations,''
Nature, {\bf 408}, 540 (2000). 

\noindent
$^5$ Moffatt, H. K., ``Moffatt replies to numismatic gyrations,''
Nature, {\bf 408}, 540 (2000). 

\noindent
$^6$ Jensen, B. L., Sumer, B. M. \& Fredsoe, J., ``Turbulent oscillatory
boundary layers at high Reynolds numbers,'' J. Fluid Mech., 
{\bf 206}, 265-297 (1989) 

\noindent
$^7$ Cutler, C. et al., ``The last three minutes: issues in
gravitational wave measurements of coalescing compact binaries,'' 
Phys. Rev. Letters, {\bf 70}, 2984-2987 (1993).

\end